# *Impairment of consciousness in Alzheimer's disease: the amyloid water-filled nanotubes manifest quantum optical coherence interfering with the normal QBD?*

Danko Dimchev Georgiev [*], Medical University Of Varna, Bulgaria

Recent discovery by *Perutz et al*. of the physical structure of the amyloid that accumulates in neurons in certain neurodegenerative diseases like Alzheimer's disease or Huntington's disease, suggests novel mechanism of consciousness impairment, different from the neuronal loss, which is the end stage of the pathogenic process. Amyloid is shown to be water-filled nanotubes made of polymerized pathologically-changed proteins. It is hypothesized that the water inside the new-formed nanotubes can manifest optical coherent laser-like excitations and superradiance similarly to the processes taking part in the normal brain microtubules as shown by *Jibu et al*. The interfering with the macroscopic quantum effects within the normal microtubules can lead to impairment of conscious experience. Experimental data in favor of quantum theory of consciousness can be obtained from the research of the amyloid nanotubes.

## 1. Quantum optical coherence in the brain microtubules

Activities within the brain's neurons are organized by dynamic scaffolding called the cytoskeleton, whose major components are microtubules. Hollow, crystalline cylinders 25 nanometers in diameter, microtubules are comprised of hexagonal lattices of proteins, known as tubulin. Microtubules are essential to cell shape, function, movement, and division. In neurons microtubules self-assemble to extend axons and dendrites and form synaptic connections, then help to maintain and regulate synaptic activity responsible for learning and cognitive functions. Microtubules interact with membrane structures mechanically by linking proteins, chemically by ions and "second-messenger" signals, and electrically by voltage fields.

While microtubules have traditionally been considered as purely structural elements, recent evidence has revealed that mechanical and quantum signaling also exist: microtubule "kinks" travel at 15 microns (2000 tubulin subunits) per second [1], microtubules vibrate (100-650 Hz) with nanometer displacement [2], microtubules optically "shimmer" when metabolically active [3], mechanical signals propagate through microtubules to cell nucleus supposing mechanism for MT regulation of gene expression [4], measured tubulin dipoles and microtubule conductivity suggest that microtubules are ferroelectric at physiological temperature [5], in the vicinity of the microtubules the water molecules show rich, ordered and systematic dynamics allowing two typical cooperative quantum phenomena called superradiance and self-induced transparency [6,7,8], tubulin states can be in superposition [9], consciousness can be result of quantum computation via applied by short laser pulses quantum gates within the microtubules in the brain cortex [10].

---

[*] *E-mail:* dankomed@yahoo.com

An essential phenomenon for the quantum brain dynamics is the quantum optical coherence in the microtubules. *Jibu et al., 1994* [6] derive the total Hamiltonian for the system of N water molecules and the quantized electromagnetic field in the region V inside the microtubule cylinder in the form

(1)
$$H = H_{EM} + \varepsilon S - \mu \sum_k (E_k^- S_k^- + S_k^+ E_k^+),$$

where μ is the electric dipole moment of a water molecule (μ =2e$_p$P with P~0.2 Å and e$_p$ - the proton charge), ε is the energy difference between the two principal energy 'eigenstates' of the water molecules (according to *Franks, 1972* it's real value is $\varepsilon \approx 200 cm^{-1}$), H$_{EM}$ is the Hamiltonian describing the quantized electromagnetic field in the region V given by

(2)
$$H_{EM} = \frac{1}{2} \int_V E^2 d^3 r,$$

E$^+$ and E$^-$ are the positive and negative frequency parts of the electric field operator given by

(3)
$$E^{\pm}(r,t) = \sum_k E_k^{\pm}(t) e^{\pm i(k \cdot r - \omega_k t)}$$

with ω$_k$ denoting the proper angular frequency of the normal mode with wave vector k, and $r^j = (x^j, y^j, z^j)$ gives the coordinates of the jth water molecule, where z denotes the axis lying along the longitudinal center axis of the microtubule cylinder (the xy-plane of the introduced Cartesian system of coordinates O$_{xyz}$ is attached parallel to one of the two ends of the microtubule, so that the origin O coincides with the center of the end cap).

$S_k^{\pm}(t)$ is collective dynamical variable for the quantized electromagnetic field in the cavity region V given by

(4)
$$S_k^{\pm}(t) = \sum_{j=1}^{N} s_{\pm}^j(t) e^{\pm i(k \cdot r^j - \omega_k t)},$$

where $s^j = \frac{1}{2}\sigma$ is the spin variable of the jth water molecule; $\sigma = (\sigma_x, \sigma_y, \sigma_z)$ and the $\sigma_x$ are Pauli spin matrices denoting the three components of the angular momentum for spin ½.

S in the total Hamiltonian is the collective dynamical variable for the water molecules given by

$$S = \sum_{j=1}^{N} s_z^j .$$

(5)

It seems worthwhile to note that the total Hamiltonian (1) is essentially of the same form as the Dicke's Hamiltonian (1954) [11] for the laser system and the Stuart-Takahashi-Umezawa Hamiltonian (1979) [12] for Quantum Brain Dynamics (QBD). Therefore, it can be expected that microtubules in the cytoskeletal structure of brain cells manifest both memory printing/recalling mechanism in QBD and laser-like coherent optical activity. The quantum dynamical system of water molecules and the quantized electromagnetic field confined inside the hollow microtubule core can manifest the specific collective dynamics called 'superradiance' by which the microtubule can transform any incoherent, i.e., thermal, and disordered molecular, electromagnetic, or atomic energy into coherent photons inside the microtubules. Analogous to superconductivity, *Jibu and Yasue, 1994* further suggest that such coherent photons created by superradiance penetrate perfectly along the internal hollow core as if the optical medium were made transparent by the propagating photons themselves. This is a quantum theoretical phenomenon called 'self-induced transparency'. That's why the coherence is not immediately lost.

*The process of superradiance can be described in four steps:*

(a) Initial state of the system of water molecules in a microtubule. Energy gain due to the thermal fluctuation of tubulins increases the number of water molecules in the first excited rotational energy state.

(b) A collective mode of the system of water molecules in rotationally excited states. A long-range coherence is achieved inside a microtubule by means of spontaneous symmetry breaking.

(c) A collective mode of the system of water molecules in rotationally excited states loses its energy collectively, and creates coherent photons in the quantized electromagnetic field inside a microtubule.

(d) Water molecules, having lost their first excited rotational energies by super-radiance, start again to gain energy from the thermal fluctuations of tubulins, and the system of water molecules recover the initial state (a) spot, and initiated movement.

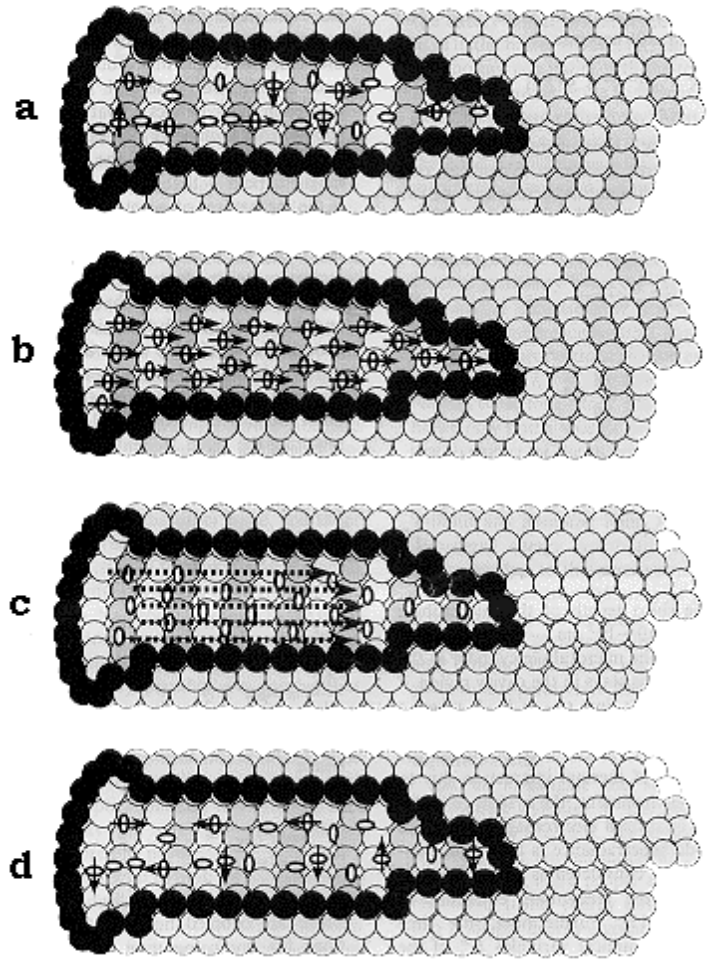

Fig. 1 | A schematic representation of the process of superradiance in a microtubule. Each oval without an arrow stands for water molecule in the lowest rotational energy state. Each oval with an arrow stands for a water molecule in the first excited rotational energy state. First a disordered ground state of water (a) in the core of the microtubule is pumped (a la Fröhlich) by disordered energy into (b) an ordered state: the "lowest rotational energy state" - the system passes into a collective dipole mode. This collective mode then (c) loses its energy collectively creating coherent photons within the microtubule... The process is cyclic (a, b, c, d, a, b), and so on. From *Jibu et al, 1994*.

## 2. Amyloid fibers are water-filled nanotubes

A recent study on amyloid fibers by *Perutz et al., 2002* [13] showed that cylindrical β-sheets are the only structures consistent with some of the x-ray and electron microscope data. Investigation of protein fibers from Alzheimer extracellular Aβ plaques (residues 11–25 fragment) via X-ray diffraction and computed reconstruction from electron micrographs revealed that all fibers give a strong 4.75-Å meridional x-ray reflection. Small angle reflections were not recorded. Optical reconstruction shows cylinders of 57 Å diameter with 37-Å-thick walls surrounding a central hole of 19.5 Å diameter. It also shows β-strands clearly stacked 4.75 Å apart with their length normal to the fiber axis. Suggested interpretation is that there are two concentric cylinders of β-sheets, the chains run normal to the fiber axis as in the reconstruction with calculated fiber diameter 60 Å. Alzheimer extracellular Aβ variant (residues 11–25 with Asp-23 → Lys substitution) revealed fibers of 35–40 Å diameter, strong 4.75-meridional and extremely weak 10-Å equatorial reflexions. The experimental findings suggest single cylindrical β-sheet made of two 14-residue peptides arranged in tandem in a circle, giving a calculated fiber diameter of 40 Å.

Based on the x-ray diffraction patterns of poly-L-glutamine, of the exon-1 peptide of huntingtin, and of a peptide corresponding to the glutamine/asparagine rich region of the yeast prion Sup35 *Perutz et al.* conclude that it forms cylinder with external diameter 30 Å and 20 glutamine residues per turn. The 3-Å-wide layer of water that adheres to the surface of proteins and is therefore unavailable as a solvent to diffusible electrolytes, then the central cylindrical cavity is only 6 Å wide, so only water and small ions can diffuse into this narrow channel.

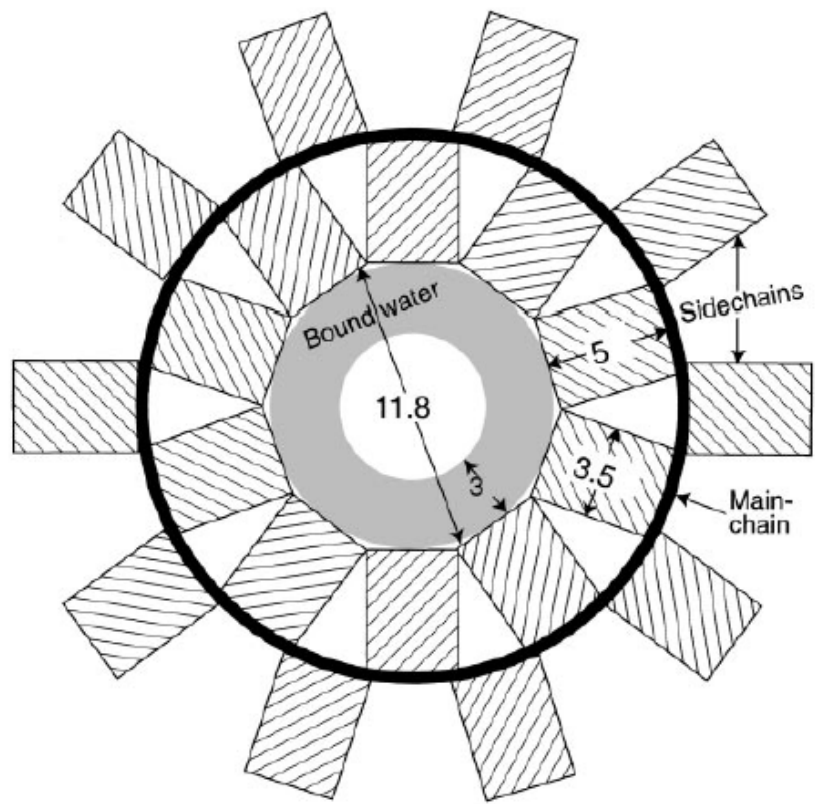

Fig.2 | Diagrammatic projection of a helical polyglutamine fiber with 20 residues per turn on a plane normal to the fiber axis. The main chain is represented by the heavy circle, and the side chains by rectangular boxes 5 Å long and 3.5 Å across. In a helix of 20 residues per turn, the terminal atoms of side chains of average length would be at the correct van der Waals distance of 3.6 Å from each other. With 18 residues per turn, this value shrinks to 3.2 Å, a little short; with 22 residues per turn, it expands to 3.9 Å, so that contact is lost. Twenty residues per turn may be the most stable structure for any sequence of residues. From *Perutz et al., 2002*.

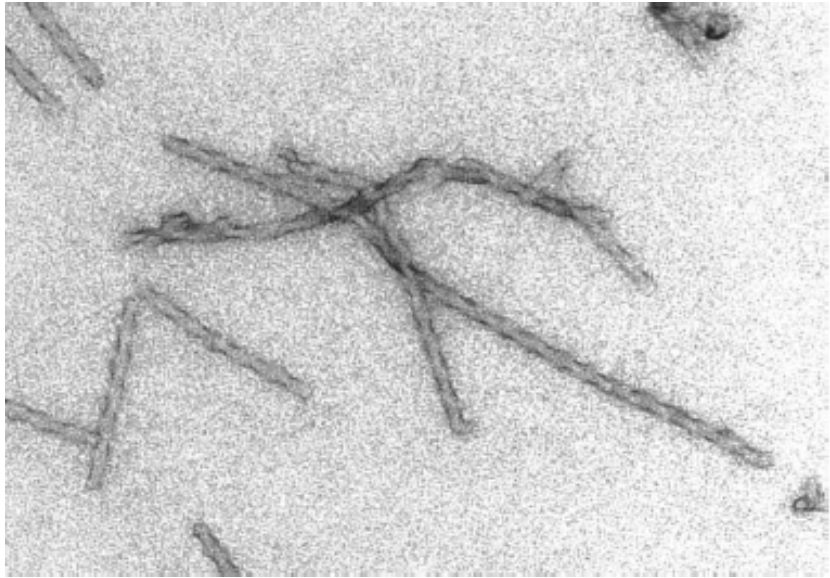

Fig. 3 | Electron micrograph of water-filled nanotube fibers made of the glutamine- and asparagine-rich region of the yeast prion Sup35. From *Perutz et al., 2002*.

## 3. Novel pathogenetic mechanism for impairment of conscious experience in AD

Alzheimer's disease (AD) is characterized by the deposition of senile plaques (SPs) and neurofibrillary tangles (NFTs) in vulnerable brain regions. SPs are composed of aggregated beta-amyloid (Aβ) peptides. Evidence implicates a central role for Aβ in the pathophysiology of AD. Mutations in βAPP and presenilin 1 (PS1) lead to elevated secretion of Aβ, especially the more amyloidogenic Aβ42. Immunohistochemical studies have also emphasized the importance of Aβ42 in initiating plaque pathology. Cell biological studies have demonstrated that Aβ is generated intracellularly. Recently, endogenous Aβ42 staining was demonstrated within cultured neurons by confocal immunofluorescence microscopy and within neurons of PS1 mutant transgenic mice. A central question about the role of Aβ in disease concerns whether extracellular Aβ deposition or intracellular Aβ accumulation initiates the disease process.

*Gouras et al., 2000* [14] report that human neurons in AD-vulnerable brain regions specifically accumulate γ-cleaved Aβ42 and suggest that this intraneuronal Aβ42 immunoreactivity appears to precede both NFT and Aβ plaque deposition. This study suggests that intracellular Aβ42 accumulation is an early event in neuronal dysfunction and that preventing intraneuronal Aβ42 aggregation may be an important therapeutic direction for the treatment of AD.

Because intraneuronal Aβ42 accumulation occurs with early AD pathology, it is possible that extracellular Aβ plaques may develop from this intraneuronally accumulating pool of Aβ42. Consistent with this possibility, *Gouras et al., 2000* have observed instances where Aβ42 appears to aggregate within the cytoplasm of neurons and where Aβ plaque staining was neuronal in shape. They have observed both diffuse plaque-like Aβ42 immunoreactivity that appears to be located directly outside neurons and early Aβ42 immunoreactivity along the axonal projections (perforant path) of early Aβ42 accumulating neurons of the entorhinal cortex and at their terminal fields, the outer molecular layer of the dentate gyrus. Brain tissue from a 64-year-old representative subject with mild cognitive impairment, stained with antibodies specific to the C-terminus of Aβ42, revealed significant amounts of region-specific intraneuronal immunoreactivity, compared with relatively little Aβ40 immunoreactivity. This intraneuronal Aβ42 staining was especially evident within pyramidal neurons of areas such as the hippocampus/entorhinal cortex, which are prone to developing early AD neuropathology.

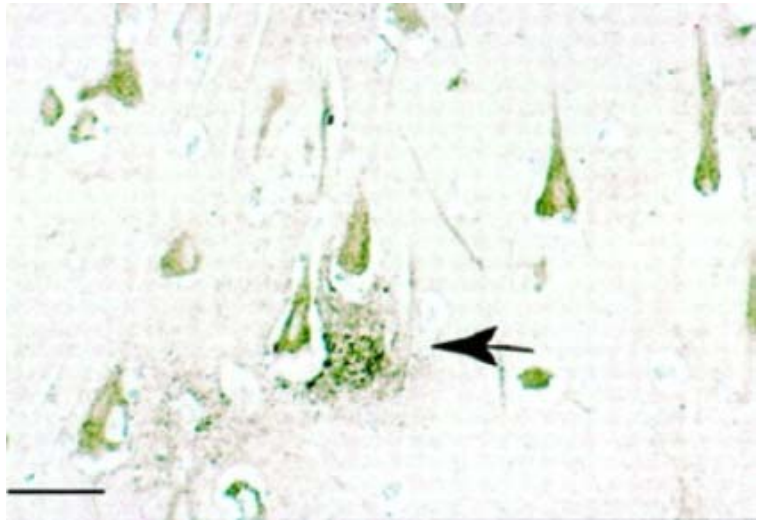

Fig. 4 | The CA1 region of a 79-year-old cognitively impaired subject demonstrates both intraneuronal Aβ42 immunoreactivity and apparent extraneuronal diffuse plaque-like staining (arrow) adjacent to a few neurons. Bar=40 μm. From *Gouras et al., 2000.*

The amyloid fibers in Alzheimer's disease may consist of two or more concentric cylindrical β-sheets or of two or more single cylindrical sheet fibers wound around each other. The complete Aβ peptide contains 42 residues, just the right number to nucleate a cylindrical shell; this finding and the many possible strong electrostatic interactions in β-sheets made of the Aβ and the absence of prolines account for the propensity of the Aβ peptide to form the amyloid plaques found in Alzheimer patients. If this interpretation is correct, amyloids consist of narrow tubes (nanotubes) with a central water-filled cavity with diameter of about 1-2 nanometers, so the water molecules in the secluded area inside the cylinder can manifest coherent excitations, just the way the normal microtubules do.

*So the consciousness disorders can be result from interfering the function (i.e. quantum computation) of the normal microtubules by the new-formed protein amyloid nanotubes, which can be well cross-linked via different filaments with the normal microtubules. After all the water molecules within the amyloid nanotubes can manifest the same laser-like coherent excitations. The neuronal loss although important is supposed to be the end stage of the pathologic process started with the intracellular accumulation of amyloid water-filled nanotube fibers.*

The other classic lesion observed in Alzheimer's original patient of 1906 is the neurofibrillary tangle. Tangles are generally intraneuronal cytoplasmic bundles of paired, helically wound 10 nm filaments (PHF), sometimes interspersed with straight filaments. Neurofibrillary tangles usually occur in large numbers in the Alzheimer brain, particularly in entorhinal cortex; hippocampus; amygdala; association cortices of the frontal, temporal, and parietal lobes; and certain subcortical nuclei that project to these regions. The subunit protein of the PHF is the microtubule-associated protein, tau. PHF are not limited to the tangles found in the cell bodies of neurons, but also occur in many of the dystrophic neurites present within and outside of the amyloid plaques. Biochemical studies reveal that the tau present in PHF comprises hyperphosphorylated, insoluble forms of this normally highly soluble cytosolic protein. The insoluble tau aggregates in the tangles are often complexed with ubiquitin. If this ubiquitination represents an attempt to remove the tau filaments by way of degradation by the proteasome, it seems to be largely unsuccessful. Phosphorylated forms of the neurofilament proteins accumulate in tangles but are not considered to be actual subunits of the PHF.

So in attempt to find some evidence in favor of the microtubule role in consciousness *Mershin, Nanopoulos and Skoulakis, 2000* [15] quote that some kind of pathology in MAP-tau proteins is the major candidate for explaining the cortical impairment and the following dementia in Alzheimer's patients.

However, the two classic lesions of AD can occur independently of each other. Tangles composed of tau aggregates that are biochemically similar to and in some cases indistinguishable from those that occur in AD have been described in a dozen or more less common neurodegenerative diseases, in which one usually finds no A$\beta$ deposits and neuritic plaques. Conversely, A$\beta$ deposits can be seen in aged, normal human brain in the virtual absence of tangles. There are also infrequent cases of AD itself that are tangle-poor, in other words very few neurofibrillary tangles are found in the neocortex despite abundant A$\beta$ plaques. It appears that in quite a few such cases, an alternate form of neuronal inclusion, the Lewy body (composed principally of alpha-synuclein protein), is found in many cortical pyramidal neurons. In other words, the Lewy body variant of AD may represent a tangle-poor form of AD that is still characterized by the usual A$\beta$ plaque formation.

*The fact that neurofibrillary tangles composed of altered, aggregated forms of tau protein occur in disorders (e.g., subacute sclerosing panencephalitis, Kuf's disease, progressive supranuclear palsy) in the absence of A$\beta$ deposition suggests that tangles can arise in the course of a variety of primary neuronal insults. As outlined by* **Dennis J. Selkoe, 2000** *the formation of tangles in AD represents one of several cytologic responses by neurons to the gradual accumulation of A$\beta$ and A$\beta$-associated proteins. Although the tangles can contribute to the impairment of consciousness they are by no means the principal way that does so, but the intracellular amyloid accumulation!*

## 4. Treatment approaches

As seen from the presentation above the Aβ seems the major pathogenetic factor in Alzheimer's disease. *Perutz et al., 2002* [13] conclude that there are any hopes for therapy. Formation of amyloids and of huntingtin aggregates is reversible. Aromatic compounds such as congo red that can insert themselves into gaps between helical turns might destabilize the cylindrical shells and initiate this process, but prevention would be more effective and probably easier to achieve.

## 5. On possible experiments that can prove/disprove quantum consciousness

The possibility that the amyloid water-filled nanotubes manifest quantum optical coherence within the neurons and interfere with the normal quantum brain dynamic can be experimentally tested. The microtubules are thought to be very sensitive to their environment, so that the quantum coherent phenomena are hard to detect *in vitro*. *Jibu et al., 1994* [6] mark that no proof of coherent photon generation or emission has been found in microtubules, but self-focused photons and self-trapped, non-thermalizing wave entities should be difficult to detect. However if amyloid nanotubes do really interfere with the quantum optical phenomena in the brain microtubules, their behavior must be manifested even *in vitro*, because both the amyloid nanotube fibers are more stable than the cellular microtubules and their mode of action is by no means controllable. For this can account the highly insulated interior of the amyloid cylinder and its smaller diameter ($d_A$ ~1-2 nm) compared to the microtubular one ($d_M$ ~15 nm). According to *Nick Mavromatos* one should take care of environmental isolation and this is why he confines the macroscopic quantum effects in smaller regions inside microtubules near the microtubular wall. In his words 25 nm diameter is too big for long lasting quantum coherent effects. In a region of 25 nm any quantum coherent state could decohere quickly within $10^{-14}$ seconds; but in thin layers *Mavromatos et al.* [16] claim you can have it up to $10^{-6}$ sec. If the macroscopic quantum phenomena *in vivo* are confined only to the region near the microtubular wall, then it will be quite hard to detect them experimentally. However, this won't be the case with the coherence inside the amyloid nanotubes, which should last orders of magnitude longer, sufficient for experimental registration.

Also in experiments trying to amplificate the number of the emitted via superradance photons, so that to be detected, it is possible to be used amyloid nanotubes instead of brain microtubules and 'harder' conditions under which the microtubules itself would be destroyed i.e. to be used the conformational stability of the amyloid fibers, which are built up mainly from β–sheets. Proving of quantum coherent optical excitations within the amyloid nanotubes will be highly supportive for analogous process taking part *in vivo* in the brain.

**Acknowledgments**: *I would like to sincerely thank to **Kunio Yasue** and **Mari Jibu** for the kindness to present me their work on quantum brain dynamics. Special thanks to **Nick Mavromatos** for the precious advises.*